\documentclass[12pt,preprint]{aastex}

\shortauthors{McCabe, Duch\^{e}ne \& Ghez}
\shorttitle{Disks in Mid-infrared Scattered Light}

\begin{document}

\title{The First Detection of Spatially Resolved Mid-Infrared Scattered Light From a Protoplanetary Disk\footnote{ The data presented herein were obtained at the W.M. Keck Observatory, which is operated as a scientific partnership among the California Institute of Technology, the University of California and the National Aeronautics and Space Administration. The Observatory was made possible by the generous financial support of the W.M. Keck Foundation.}}
\author{C. McCabe, G. Duch\^{e}ne \& A.M. Ghez}
\affil{Department of Physics and Astronomy, University of California, 
Los Angeles}
\authoraddr{405 Hilgard Ave, Los Angeles, CA 90095-1562}
\email{mccabe@astro.ucla.edu, duchene@astro.ucla.edu, ghez@astro.ucla.edu}

\begin{abstract}
We report spatially resolved 11.8\,\micron\, images, obtained at the
W. M. Keck 10\,m telescope, of the protoplanetary disk around the
pre--main-sequence star HK~Tau B. The mid-infrared morphology and
astrometry of HK~Tau B with respect to HK~Tau A indicate that the flux
observed in the mid-infrared from HK~Tau B has been scattered off the
upper surface of its nearly edge-on disk. This is the first example of
a protoplanetary disk observed in scattered light at mid-infrared
wavelengths. Monte Carlo simulations of this disk show that the extent
(FWHM $\sim$0\farcs5, or $\sim$70 AU) of the scattered light nebula in
the mid-infrared is very sensitive to the dust size distribution. The
11.8\,\micron\, measurement can be best modelled by a dust grain
population that contains grains on the order of 1.5-3\,\micron\, in
size; grain populations with exclusively sub-micron grain sizes or power
law size distributions that extend beyond 5\,\micron\, cannot
reproduce the observed morphology. These grains are significantly
larger than those expected in the ISM implying that grain growth has
occurred; whether this growth is a result of dust evolution within the
disk itself or had originally occurred within the dark cloud remains
an open question.

\end{abstract}


\section{Introduction}
The first step toward planet formation is thought to be the growth of
dust grains in protoplanetary disks surrounding young stellar
objects. Models suggest that by the T Tauri stage of evolution ($\sim$
1 Myr), significant grain growth should have already occurred, with
the presence of large bodies expected (e.g., Beckwith, Henning \&
Nakagawa 2000). Observationally, this has been hard to prove
conclusively. Although tentative evidence for grain growth was first
suggested in measurements of the thermal emission from these young
disks at millimeter wavelengths, grain growth is only one of a handful
of possible explanations for the observed flattening of the wavelength
dependence of opacity (e.g., Beckwith \& Sargent 1991). In principle,
the effect of grain growth should be more marked in long wavelength
scattered light than in thermal emission, since scattered light is
most sensitive to grain sizes on order of the wavelength of
observation.  The longest wavelength, however, at which resolved
scattered light from T Tauri disks has been detected is 2.2\,\micron\,
(e.g., Cotera et al. 2001; McCabe, Duch\^{e}ne \& Ghez 2002), which
only effectively probes grain sizes up to the maximum grain size
expected in most model ISM distributions ($\sim$ 1\,\micron\,; Kim,
Martin \& Hendry 1994). These experiments, therefore, have not yet
been tremendously sensitive to grain growth. At mid-infrared
wavelengths, scattered light experiments probe grain sizes roughly an
order of magnitude larger than the standard ISM cut-off, as has been
successfully demonstrated in observations of a more evolved
post-main-sequence nebula (Tuthill et al. 2002). Such experiments
applied to the pre-main-sequence stage of evolution potentially offer
a powerful way to probe the initial stages of grain growth.

The HK~Tau binary system is an ideal candidate for the detection of
scattered light at mid-infrared wavelengths. While the primary,
HK\,Tau A, has yet to be spatially resolved at any wavelength, the
secondary, HK~Tau B, is detected as an extended nebula at optical
(Stapelfeldt et al. 1998), near-infrared (Stapelfeldt et al. 1998;
Koresko 1998), and millimeter wavelengths (Duch\^{e}ne et
al. 2003). The near-infrared and optical scattered light images are
best modelled as light being scattered off the flared outer edges of
the upper and lower surfaces of a nearly edge-on, optically thick
disk, whereas the millimeter observations probe the dust thermal
emission. While there are no existing reports of the detection of
spatially resolved mid-infrared scattered light from any
protoplanetary disk or debris disk, the advent of 10 meter class
telescopes should permit such a detection for HK~Tau B.  The disk is
large ($\sim$1\farcs5, Stapelfeldt et al. 1998) compared to the
mid-infrared diffraction limit of a 10~m telescope ($\sim$0\farcs2 at
10\,\micron), and the measured lower limit to the line of sight
extinction within the disk (Stapelfeldt et al. 1998) suggests that it
is optically thick at mid-infrared wavelengths ($\tau_{11.5 \micron}>
3.7$, using the ISM extinction law of Rieke \& Lebofsky 1985). As
such, any competing thermal emission arising in the hot inner disk
region is removed from the line of sight.  An additional advantage of
the HK~Tau system is that HK~Tau A, located only 2\farcs4 away from
HK~Tau B, provides an instantaneous measurement of the point spread
function and an astrometric reference point.

Models of the dust properties of the HK~Tau B disk in both the
scattering and thermal regime have led to contradictory results
(Stapelfeldt et al. 1998; D'Alessio et al. 2001; Duch\^{e}ne et
al. 2003), with evidence for and against the presence of grain
growth. While the optical and near-infrared scattered light images are
consistent with standard ISM dust size distributions, D'Alessio et
al. (2001), using SED fitting constrained by the outer radius of the
disk, suggest a grain size distribution that extends up to at least 1
mm. However, Duch\^{e}ne et al. (2003), in resolving the 1.3
millimeter flux between the two components of the system, limit the
maximum grain size to 200\,\micron, ruling out the presence of
millimeter sized particles as predicted by D'Alessio et al. Furthermore,
Duch\^{e}ne et al.  find that the thermal millimeter and optical
scattered light results are hard to reconcile with a single population
of dust grains, leading them to suggest that HK~Tau B has a layered
disk structure, with the surface layer showing no evidence for
1--3\,\micron\, sized grains, assuming a power law size distribution.
This latter conclusion can be readily tested; scattered light images
at $\sim$10\,\micron\, are by far the most direct probe of the
presence of supra-micron sized grains in the surface layer of the
disk.

In this paper, we report observations of HK~Tau using the Keck 10~m
telescope which resolve the disk at 11.8\,\micron. The observations
and data reduction are outlined in \S2 and the resulting disk and
binary parameters are investigated in \S3. We argue that the observed
mid-infrared flux is actually scattered light, and, comparing the
observations to Monte Carlo scattering simulations, place constraints
on the size distribution in \S4.

\section{Observations}
On 2002 November 13 (UT), observations of HK~Tau ($\alpha = 04^{h}
28^{m} 48.^{s}85\, \delta = 24\arcdeg\, 17\arcmin\, 56\farcs2$, B1950)
were made on the W.M. Keck I telescope with the Long Wavelength
Spectrometer ({\it LWS}, Jones \& Puetter 1993) in imaging mode
through the SiC filter ($\lambda_{o}$ = 11.78\,\micron\,,
$\Delta\lambda$ = 2.34\,\micron). HK~Tau was observed in the standard
chop-nod mode, with the chopper running at a frequency of 5 Hz and an
amplitude of 10\arcsec, which moved the object out of the field of
view in the chop beam. We chopped along a position angle of
60$^{\circ}$, which was chosen to avoid chopping along either the
binary position angle or the edge-on disk axis. The chop-nod cycle was
repeated 13 times, resulting in a total on-source integration time of
312 seconds. The photometric standard $\alpha$ Ari was observed
regularly throughout the night, establishing that atmospheric
conditions were photometric.

Each chop-nod cycle is combined using the standard `double difference'
technique designed to remove the sky and telescope thermal signatures
from the data. The final image is the median of 12 of the double
differences, shifted using HK~Tau A, which is clearly detected in each
frame, as a reference point. One frame suffered from chop smearing and
was therefore not included in the analysis. A similar procedure is
used to analyze $\alpha$ Ari and 6 known binary systems in Taurus
observed as part of a separate study (McCabe et al. in preparation);
the binary stars provide estimates of the {\it LWS} plate
scale and orientation, measured to be 0\farcs081 $\pm$ 0\farcs002 and
-3\fdg6 $\pm 1^{\circ}$, respectively.\footnote{The chosen binaries
range in separation from 2-6\arcsec. At the smallest separation,
orbital motion from a circular orbit over a 10 year period would
correspond to at most a 0\fdg78 change in position angle at the
distance to Taurus (d=139 pc).  This analysis is therefore unaffected
by any potential orbital motion.} The resulting point spread function
of these observations are nearly diffraction limited (FWHM $\simeq$
0\farcs29).

The absolute photometric calibration for the new SiC filter is
established using calibrated stellar spectra of $\alpha$ Ari and
$\alpha$ Lyr, which were observed along with $\alpha$ Ari on the
previous night (Cohen et al. 1992; Cohen et al. 1999). The total flux
density expected from these two observed standards is modelled by
integrating the calibrated spectra through the transmission profiles
of the filter, the model atmosphere above Mauna Kea assuming 1.5
airmasses and 1.6 mm of precipitable water\footnote{These ATRANS
atmospheric transmission models, with varying values of precipitable
H2O and airmass, are available on the Gemini Observatory website,\\
http://www.gemini.edu/sciops/ObsProcess/obsConstraints/ocTransSpectra.html}
(Lord 1992), and the quantum efficiency of the {\it LWS} detector.
The modelled flux densities for $\alpha$ Lyr and $\alpha$ Ari are 28.3
Jy and 58.6 Jy respectively. The difference between the modelled and
observed flux density ratio between these two stars is 6\%, which we
use as an estimate of the uncertainty in this calibration process.

\section{Results}

Figure 1 shows the final 11.8\,\micron\, image, in which two sources
are detected with a signal to noise ratio of 63 and 8 respectively.
Photometry is carried out on both components using a 0\farcs49 radius
circular aperture, resulting in a flux density ratio of 30 $\pm$ 4 and
an absolute flux density of 208 $\pm$ 13 mJy and 6.9 $\pm$ 1.0 mJy for
the primary and secondary, respectively. A $\chi^{2}$ fit of a 2
dimensional gaussian to each component, using the {\it n2gaussfit}
task in IRAF\footnote{IRAF is distributed by the National Optical
Astronomy Observatories, which is operated by the Association of
Universities for Research in Astronomy, Inc., under contract to the
National Science Foundation.}, is done to provide relative astrometry.
Separated by 2\farcs228 $\pm$ 0\farcs056 at a position angle of
170.3\degr $\pm$ 1.2\degr, the two sources have relative positions in
the mid-infrared that are consistent with those of HK~Tau A and B at
shorter wavelengths (e.g., Leinert et al. 1993; Simon et al. 1995); we
therefore conclude that we are detecting mid-infrared emission from
both components of the HK~Tau binary system.

The most striking aspect of Figure 1 is that the secondary, HK~Tau B,
is spatially extended. The best $\chi^{2}$ gaussian fit
($\chi^{2}_{dof}$=0.91 over 0.65 square arcseconds) finds a FWHM of
0\farcs50 $\pm$ 0\farcs05 aligned along a position angle of 51\degr
$\pm$ 6\degr. Using the position of HK~Tau A, along with the relative
plate scale and orientation of the cameras, the mid-infrared image of
HK~Tau is aligned with the optical WFPC2 image (Stapelfeldt et
al. 1998) in the zoomed-in panel in Figure 1. The uncertainty in each
direction is 0\farcs047; the WFPC2 platescale and orientation
uncertainties are negligible compared to those of {\it LWS} (Holtzman
et al. 1995), therefore the positional uncertainty is dominated by the
uncertainty in the {\it LWS} platescale and orientation measurements.
Comparison with the WFPC2 observations show that the extended
mid-infrared emission is aligned along the same position angle of the
scattered light disk seen in the visible (51\degr $\pm$ 6\degr\,
vs. 40\degr) and is coincident with the northwest nebula to within
0\farcs03 (0.6$\sigma$) in the direction perpendicular to the disk.
In contrast, the offset between the disk midplane, represented by the
center of the dark lane in the WFPC2 images, and the extended
mid-infrared flux is significant (0\farcs146; 3.1$\sigma$).

The 11.8\,\micron\, flux appears to be slightly shifted from the peak
of the visible scattered light disk (see Figure 1). Comparing the disk
centers from the gaussian fitting we find an offset of 0\farcs105
(2.2$\sigma$) along the disk axis.  Deeper observations of this system
are needed to determine whether this asymmetry is real.  The presence
of an asymmetry would not be all that unexpected; a number of
protoplanetary disks show evidence for some level of scattered light
asymmetry. For example, both GG Tau (Silber et al. 2000; McCabe,
Duchene \& Ghez 2002) and HH 30 (e.g., Burrows et al. 1996) show
structure, and HK~Tau B itself shows evidence for a non-symmetric
scattered light morphology in the southeastern nebula (Stapelfeldt et
al. 1998).

\section{Discussion \& Conclusions}

\subsection{Nature of the Emission Process}

The nature of HK~Tau B's mid-infrared flux can be established on the
basis of the observed morphology.  By far the majority of mid-infrared
photons from a typical T Tauri star+disk system (e.g., $>$90\% in the
models of Lachaume et al. 2003, and private communication) are emitted
thermally from the central 1 AU region of the disk (1\,AU = 7\,mas at
a distance of 139 pc to Taurus; Bertout, Robichon \& Arenou 1999). The
observed mid-infrared morphology of this edge-on disk therefore
depends primarily on the optical depth along the line of sight through
the disk. If the disk is optically thin, the observer will see an
unresolved point source positioned close to the midplane of the disk.
However, if the disk is optically thick, the mid-infrared emission
from this region will be totally absorbed, allowing the detection of
scattered light from the flared outer edges of the upper and lower
surfaces of the disk, a case which applies to the optical and
near-infrared images of HK~Tau B. \footnote{Thermal emission from the
outer edge of the disk is expected to be negligible at these
wavelengths as the dust temperature in these regions is $\sim$10\,K
(Stapelfeldt et al. 1998; D'Alessio et al. 2001).} In this scenario,
the mid-infrared flux will be aligned with the scattered light nebulae
seen in the optical and near-infrared and may possibly be spatially
resolved, depending on the scattering properties of the dust. The
observed mid-infrared flux is coincident with the northwest nebula, instead of
the disk midplane and is also extended along the same position
angle (\S3). This indicates that the observed mid-infrared flux is produced
by thermally emitted photons originating from the inner disk region
being scattered off the flared outer edge of the disk.

\subsection{Dust Grain Properties}

The morphology of the scattered light from the disk is expected to be
highly dependent on the properties of the dust population present in
the disk.  Typical ISM dust models produce highly isotropic scattering
in the mid-infrared regime (see, for example, Figures 1 and 2 of
Wolfire \& Churchwell 1994). As such, to first order one would expect
the entire length of each nebula to be seen in scattered light.
Alternatively, if the disk contains large grains ($ 2\pi a \gtrsim
\lambda$), which are highly forward-throwing, we would see an almost
unresolved point-like source in the center of the northwest nebula.
The observed scattered light disk morphology is between these two
extremes, being clearly resolved but not illuminated across its full
extent.  This suggests that scattering at 11.8\,\micron\,, while not as
forward-throwing as that seen in the visible, is not as isotropic as
that expected from ISM dust grains either.

We performed Monte Carlo scattering simulations of the disk in order
to confirm this qualitative behaviour and to place quantitative limits
on the size of grains responsible for the scattered light. The
simulations, which are similar to those used in McCabe, Duch\^{e}ne \&
Ghez (2002), employ Mathis \& Whiffen (1989) dust properties in an
$n(a) \propto a^{-3.7}$ power law grain size distribution. We fix
$a_{min}$, the minimum grain size in the distribution, to be 0.1
\micron\, and vary the maximum grain size to see how sensitive the
scattered light distribution is to the presence of large grains.
While $a_{min}$ is larger than normally used in a standard ISM grain
size distribution, because we are working in the mid-infrared the
scattering properties are not at all sensitive to smaller grains, and
therefore the simulations are unaffected by this choice. The best
fitting disk model for HK~Tau B from Stapelfeldt et al. (1998) is used
to define the geometry of the disk. The simulations confirm that the
scattered light distribution along the northwest nebula, through the
value of the dust asymmetry parameter, $g$, is very sensitive to the
upper limit of the grain size distribution.  As expected, models with
exclusively sub-micron sized grains yield nebulae with FWHM greater
than 1\arcsec, while models with a$_{max} \ge$ 5\,\micron\, result in
an almost unresolved point-like source in the center of the northwest
nebula. Models sampling a wide range of $g$ values show that the 11.8
\micron\, data are fairly well represented by models with
$g=0.4-0.55$. We have assumed here that the disk is symmetric about
the minor axis. The effect of the possible asymmetry can be
qualitatively estimated by taking the HWHM along each half of the disk
axis from the center of the disk and doubling it to find the potential
range in FWHM. We find a range of FWHM of 0\farcs3--0\farcs7, which
increases the range of best fit $g$ values to 0.15--0.83.

For a single size dust population, $g$ values of 0.4--0.55 at 11.8\,\micron\,
correspond to grain sizes in the range 2.5--3.2\,\micron\,, implying
that grains in this size range must exist in the surface of the
HK\,Tau B disk. Including the possibility of a disk asymmetry (with
$g$ ranging from 0.15--0.83) reduces the minimum grain size
from 2.5\,\micron\, to $\sim$1.5\,\micron. This does not suggest that
the mean grain size is this large, but that the 11.8\,\micron\,
scattering cross-section-weighted mean size is.  Only a very small
amount of such large grains is needed to achieve this, but these
results show that grains of this size must be present in the disk.
These grains are significantly larger than those expected in the ISM,
implying that grain growth has occurred in the upper surface of this
disk. Such large grains are not excluded from extinction fits to dense
cloud regions (e.g., $R_{V} >$ 4, Weingartner \& Draine 2001),
therefore it remains an open question as to whether the large grains
are a result of dust evolution within the disk or were already present
in the dense cloud from which this system formed.

It should be noted that a single ISM-like power law size distribution
extending to $\sim$2\,\micron\, produces $g$ values in the optical
that are significantly higher than the best-fit $g$ found by
Stapelfeldt et al. (1998), a model contradiction that has been pointed
out by Duch\^{e}ne et al. (2003).  One potential way to reconcile the
various wavelength constraints is a more complex size distribution,
such as a power law with a gradual fall-off. Such a modified size
distribution has already been suggested for ISM grains (e.g., Kim,
Martin \& Hendry 1994; Weingartner \& Draine 2001). The results from
observations of this disk at multiple wavelengths also clearly show
that analysis of dust scattering from a single wavelength can provide
misleading results. A more thorough characterization of the grain
properties and disk geometry will require simultaneous modelling of
deep images over as wide a range of wavelengths as possible. This
study, however, unambiguously shows the need for 1--3\,\micron\,
grains to be included in any dust model of this system.\\

The authors thank Randy Campbell and Ron Quick at the Keck Observatory
for their help in obtaining these observations and Karl Stapelfeldt
and Eric Becklin for stimulating discussions. We also thank the
anonymous referee whose excellent comments have added to this
work. Support for this work was provided by the Packard
Foundation. The authors also wish to recognize and acknowledge the
very significant cultural role and reverence that the summit of Mauna
Kea has always had within the indigenous Hawaiian community.  We are
most fortunate to have the opportunity to conduct observations from
this mountain.

\clearpage
\figcaption[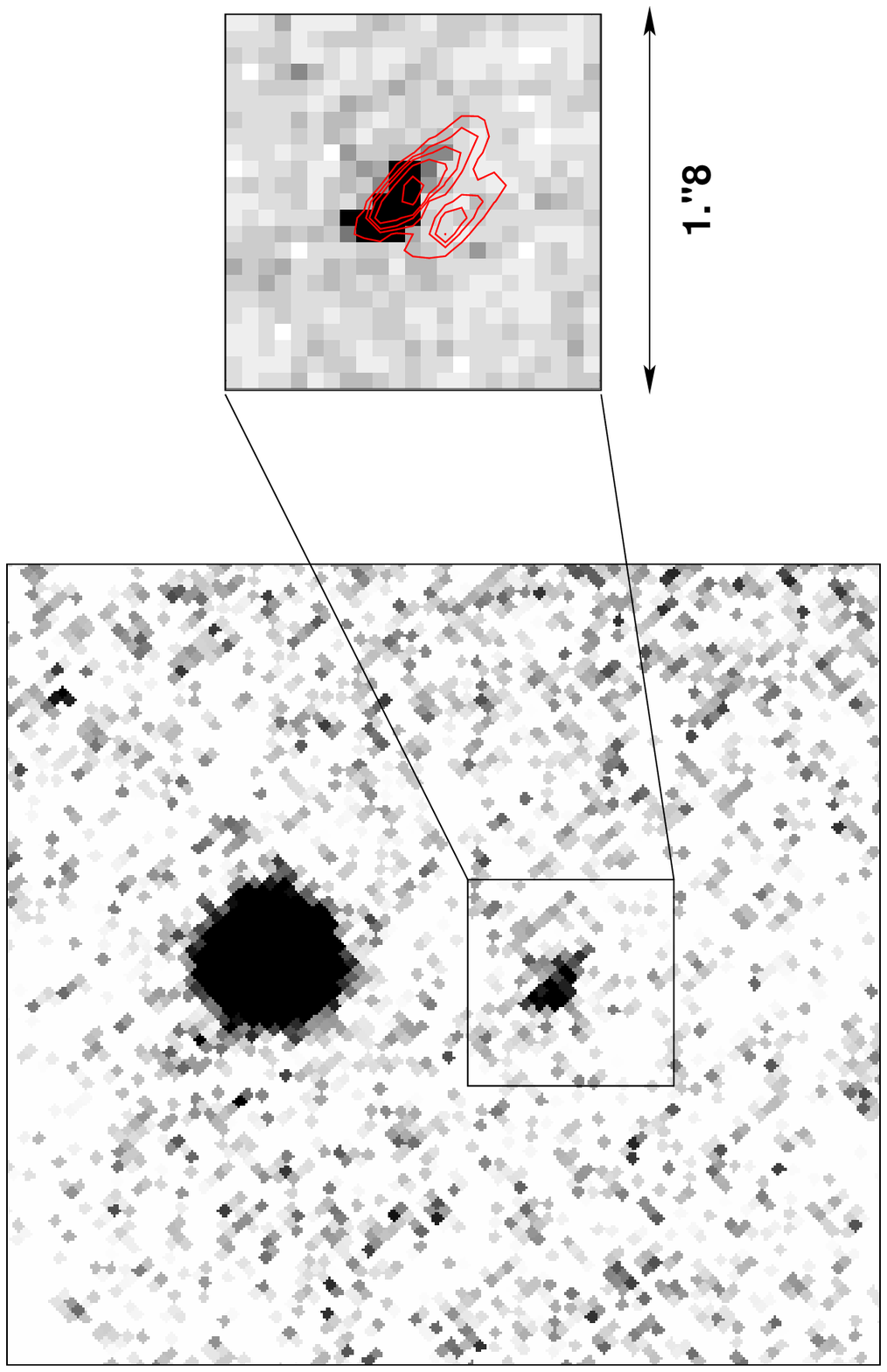]{The final image of HK~Tau at 11.8 \micron\,. The
zoomed-in panel shows the 11.8\,\micron\, disk in grayscale overlaid
with a contour plot of the WFPC2/HST image (Stapelfeldt et al. 1998)
which has been re-sampled at the {\it LWS} platescale; the contours are set
at 7, 14, 20, 30, 65\% of the peak disk flux seen in the
visible. These images are aligned using HK~Tau A, and oriented north
up, east to the left.  \label{hkcontour}}

\begin{figure}
\figurenum{1}
\plotone{f1.ps}
\end{figure}


\begin{thebibliography}{}
\bibitem[Beckwith \& Sargent (1991)]{BS91} Beckwith, S.V.W. \& Sargent, A.I., 1991,ApJ, 381, 250 
\bibitem[Beckwith, Henning \& Nakagawa (2000)]{BHN} Beckwith, S.V.W., Henning, T., \& Nakagawa, Y., 2000, in Protostars and Planets IV, eds Mannings, Boss \& Russel, pg 533
\bibitem[Bertout, Robichon \& Arenou (1999)]{BRA} Bertout, C., Robichon, N., \& Arenou, F., 1999, A\&A, 352, 574 
\bibitem[Burrows et al. (1996)]{B96} Burrows, A., Stapelfeldt, K.R., Watson, A.M., Krist, J.E. et al. 1996, ApJ, 473, 437
\bibitem[Cohen et al. (1992)]{C92}Cohen, M., Walker, R.G., Barlow, M.B., \& Deacon, J.R. 1992, AJ, 104, 1650 
\bibitem[Cohen et al. (1999)]{C99}Cohen, M., Walker, R.G. Carter, B., Hammersley, P., Kidger, M \& Noguchi, K. 1999, AJ, 117, 1864
\bibitem[Cotera et al. (2001)]{C01} Cotera, A., Whitney, B., Young E., Wolff, M.J., Wood, K., Povich, M., Schneider, G., Rieke, M., \& Thompson, R., 2001, ApJ, 556, 958 
\bibitem[D'Alessio et al. (2001)]{DACH} D'Alessio, P., Calvet, N., \& Hartmann, L., 2001, ApJ, 553, 321
\bibitem[Duch\^{e}ne et al. (2003)]{D03} Duch\^{e}ne, G., M\'{e}nard, F., Stapelfeldt, K \& Duvert, G. 2003, A\&A, 400, 559, astroph/0212512
\bibitem[Holtzman et al. (1995)]{H95} Holtzman, J., Hester, J.J., Casertano, S., Trauger, J.T. et al. 1995, PASP, 107, 156
\bibitem[Jones \& Puetter (1993)]{lws} Jones, B., \& Puetter, R., 1993, Proc. SPIE, 1946, 610
\bibitem[Kim, Martin \& Hendry (1994)]{KMH}Kim, S., Martin, P.G., \& Hendry, P.D., 
1994, \apj, 422, 164  
\bibitem[Koresko (1998)]{K98}Koresko, C.D., 1998, ApJ, 507, L145
\bibitem[Lachaume et al. (2003)]{329}Lachaume, R., Malbet, F. \& Monin, J-L., 2003, A\&A in press, astroph/0206307
\bibitem[Leinert et al. (1993)]{L93}Leinert, C., Zinnecker, H., Weitzel, N., Christou, J., Ridgway, S.T., Jameson, R., Haas, M. \& Lenzen, R., 1993, \aap, 278, 129
\bibitem[Lord (1992)]{Lord}Lord, S.D. 1992, NASA Technical Memor. 103957
%\bibitem[Mannings \& Emerson (1994)]{ME94}Mannings, V. \& Emerson, J.P., 1994, MNRAS, 267, 361
\bibitem[Mathis \& Whiffen (1989)]{MW}Mathis, J.S. \& Whiffen, G., 1989, ApJ, 341, 808
\bibitem[McCabe, Duch\^{e}ne \& Ghez (2002)]{MGD} McCabe, C., Duch\^{e}ne, G., \& Ghez, A.M., 2002, ApJ, 575, 974
\bibitem[Rieke \& Lebofsky (1985)]{RL} Rieke, G. \& Lebofsky, M.J., 1985, ApJ, 288, 618
\bibitem[Silber et al. (2000)]{Joel2000} Silber, J., Gledhill, T., Duch\^{e}ne, G. \& M\'{e}nard, F., 2000, ApJ, 536, L89
\bibitem[Simon et al. (1995)]{S95}Simon, M., Ghez, A. M., Leinert, Ch., Cassar, L., Chen, W. P., 
Howell, R. R., Jameson, R. F., Matthews, K., Neugebauer, G., Richichi, A., 1995, ApJ, 443, 625
\bibitem[Stapelfeldt et al. (1998)]{wfpcdisk} Stapelfeldt, K.R, Krist, J.E., M\'{e}nard, F. \& Bouvier, J., 1998, ApJ, 502, L65
\bibitem[Tuthill et al. (2002)]{T92} Tuthill, P., Men'shchikov, A.B., Schertl, D., Monnier, J.D., Danchi, W.C. \& Weigelt, G., 2002, A\&A, 389, 889
\bibitem[Weingartner \& Draine (2001)]{WD01}Weingartner \& Draine, 2001, ApJ, 548, 296
\bibitem[Wolfire \& Churchwell (1994)]{WC94}Wolfire \& Churchwell 1994, ApJ, 427, 889

\end{thebibliography}
\end{document}